\documentclass[10pt,letterpaper]{article}
\usepackage[top=0.85in,left=2.75in,footskip=0.75in]{geometry}

\usepackage{amsmath,amssymb}

\usepackage{changepage}

\usepackage{textcomp,marvosym}

\usepackage{cite}

\usepackage{nameref,hyperref}

\usepackage[nopatch=eqnum]{microtype}
\DisableLigatures[f]{encoding = *, family = * }

\usepackage[table]{xcolor}

\usepackage{array}

\newcolumntype{+}{!{\vrule width 2pt}}

\newlength\savedwidth

\raggedright
\usepackage[aboveskip=1pt,labelfont=bf,labelsep=period,justification=raggedright,singlelinecheck=off]{caption}

\makeatletter
\renewcommand{\@biblabel}[1]{\quad#1.}
\makeatother

\usepackage{lastpage,fancyhdr,graphicx}
\usepackage{epstopdf}
\fancyheadoffset[L]{2.25in}
\fancyfootoffset[L]{2.25in}
\begin{document}
\vspace*{0.2in}

\begin{flushleft}
{\Large
\textbf\newline{A computational model of spatial politics: {\color{black} Hotelling-Downs} model as statistical physics} 
}
\newline
\\
Christopher Campbell\textsuperscript{1},
Graeme J. Ackland\textsuperscript{1*}
\\
\bigskip
\textbf{1} School of Physics and Astronomy, University of Edinburgh, Edinburgh, Scotland, UK

\bigskip

%
%





* gjackland@ed.ac.uk

\end{flushleft}
\section*{Abstract}
The Hotelling-Downs model considers parties changing policy to maximise their vote-share.   Where policy position lies on a left-right axis, it describes a tendency for political parties to move towards centrist platforms.  This is in contrast with widely observed  political polarisation.
We extend the model to two dimensions, with many parties and with single and multiple-peaked voter distribution.   We find that a two party system reduces polarisation,  even if voters are polarised with a  bimodal distribution.  By contrast, multiparty systems induce polarisation, even when most voters favour moderate position. 
We model the effect of turnout and activists as influences on the parties, showing that this results in more polarisation, even in a two-party system.   This suggests that polarisation of parties can be driven by abstention, intra-party politics and turnout on the extremes.  
 In the two-party case, the winning  party's positions are  more moderate than the views of their supporters but better representative of the electorate as a whole. With polarisation, individual voters are better able to find a party which represents their views, but the government (winning part or coalition) is less representative of the population, even when the population has a clear consensus on all issues. 


\section*{Introduction}
Party polarisation has presented a problem for political theorists despite its prevalence in nations across the globe. While it is common for competing parties to seek the centre-ground as well as for them to diverge, there is still no unified model which can explain both behaviours. This work demonstrates a new Downsian model which can support both convergence and divergence, and which might serve as a basis for more complete models in future.

The spatial model was first introduced to economics by Hotelling in his `Stability in Competition'\cite{Hotelling}. Downs applied this model to voters and political parties in his seminal `An Economic Theory of Political Action in a Democracy'\cite{Downs}. The main result of this spatial model of politics was to predict party convergence around the median voter (Hotelling's law). It was subsequently made more robust\cite{calvert_robustness_1985}. This prediction was very successful for a time when the two major US parties appeared to advocate nearly identical policies. However, in the face of significant party polarisation in the US and elsewhere over the next decades, the Hotelling-Downs model no longer seems to accurately reflect reality\cite{layman_party_2006}. Boxell \emph{et al.} have found increasing polarisation in numerous countries, with elite polarisation and affective polarisation strongly correlated \cite{BoxellCountryTrends}. 
Recently, political parties have focussed on their base more than swing voters.  For example, members of the US Congress have catered to extreme voters in their state and the US parties have become more polarised in multiple independent opinion dimensions\cite{fiorina1999, poole_polarization_1984, layman_party_2002,young2026new}. Fiorina and Abrams argue that the evidence is much less clear for polarisation among voters than among elites, but agree that significant party sorting has occurred\cite{Fiorina-PolarisationTypes}. Roughly speaking, affective polarisation refers to how much followers of one party dislike followers of another, and elite polarisation refers to divergence among elites including legislators. Party sorting is the tendency for voters to agree with one party on all issues rather than only on some, though they are not necessarily more extreme. This trend for voters to agree with a party's whole set of views, its ideology, makes it reasonable to utilise a low-dimensional opinion space, {\color{black} Two dimensions are sufficient to describe the case where there is a two-peaked distribution in opinion space, despite a consensus position on all single-issues.  }

Political polarization is the divergence of political attitudes to ideological extremes, where opinions split into opposing, antagonistic camps. In the Downsian context, polarisation may mean two things:  either, a very broad range of held views, compared to the range viewed as acceptable by individuals; or, a distribution of views with  peaks away from the centre.

Duverger's law {\color{black} - that so-called "first past the post" systems lead to two-party politics - } has been shown to be consistent with polarisation in a spatial model accounting for turnout and the threat of third-party candidates\cite{Callander&Wilson-Duverger}. 
Some models do yield non-convergent equilibria, but require complete turnout\cite{besley_economic_1997, osborne_model_1996, wittman_candidate_1983}. Divergence in a two-party system can be achieved using the threat of third-party entry provided the third party can enter arbitrarily close to an existing party\cite{palfrey_spatial_1984}. This, however, is unsatisfactory since new parties are generally formed far from established ones. 
Other authors have questioned the assumptions in Downsian models, for instance the nature of the space in which parties move\cite{Stokes1963, Grofman2004}. Other significant advancements to theory were made in the landmark `A Theory of the Calculus of Voting'\cite{calculus-of-voting}.
A rigorous game-theoretic treatment of Hotelling-Downs problems and their computability is given by Harrenstein \emph{et al.}\cite{harrenstein2021a}. Ours is not the first computational model applied to the field, nor the first based on methods in natural sciences\cite{laver2012,social_stat_phys, galam_application_1999}. Laver and Sergenti have presented such a model using NetLogo with party actions reminiscent of foraging strategies in ecology with some success\cite{laver2012}. Axelrod \emph{et al.} have also employed an agent-based model based on the opinion dynamics of the voters themselves\cite{axelrod}, rather than party dynamics.
Magyar \emph{et al.} have reviewed recent attempts to include other considerations such as valence issues and note that some spatial models have produced polarisation, such as when accounting for imperfect turnout and nonpolicy characteristics\cite{DownsianReview, AdamsMerril2003-TurnoutDiv}. They find that parties are motivated by duelling centripetal and centrifugal forces, drawing the parties to centrist and polarised positions respectively, though the details of these forces and their relationship to valence is disputed. Furthermore, they note that models with more than two parties usually fail to produce stable equilibria, and that the same is true for those `scarce' and `complicated' models involving more than one policy dimension. This article presents such a multi-party, multi-dimensional model capable of supporting both polarised and consensual equilibria.

The term "equilibrium" is used for least three different meaning.  The "Nash Equilibrium" is when no individual party can change policy to improve its position, allowing for indefinitely large changes.  "Static equilibrium" means that no {\it infinitesimal} moves, including coordinated multi-party changes, can improve party positions.  "Dynamic Equilibrium" - normal in statistical physics - describes a state where parties move continuously in a restricted region of opinion state.

In addition to the turnout effect,  another behaviour is what we call activists, which represents the influence of extreme political actors and the fact that people with more extreme opinions will preferentially become candidates and participate in elections. Activists typically manifest as a centrifugal force, though they can also be centripetal, for example to model a charismatic moderate leader.

The primary results  are as follows: we confirm the findings of some previous models that the Downsian spatial framework can be adapted to produce equilibria with party polarisation, and we present a new kind of model reliant on numerical simulation and mathematical tools common in the natural sciences to easily simulate the movement of political parties under different forces. In particular, we find a negative correlation between turnout and polarisation both theoretically and empirically.
In the next section we describe our model, with the mathematical details of our implementation in \nameref{model}. We then report our results, followed by a discussion about the relationship to empirical evidence, previous theories and conclusions.

The simulation code is available at \url{https://github.com/CC-sci/parties-in-opinion-space}, and an online demo at \url{https://www2.ph.ed.ac.uk/~gja/opinion.html}.  A gamified version of the 1D model is also available \url{https://www2.ph.ed.ac.uk/~gja/hotelling_game.html}

\section*{Model}
The classic work of Downs is founded on the assumption that parties act to maximise their votes\cite{Downs}.
Here we extend Downs' idea that parties maximize votes, noting how it is similar to the thermodynamic idea of particles minimizing energy. Using principles of statistical physics, akin to Hotelling-Downs, we explore both one-dimensional and two-dimensional opinion spaces\cite{Stokes1963}, represented by $ \vec{x}=(x,y) $ with $x,y$ representing voter opinion or party policy on two issues.

Voter distributions are considered single- or double-peaked, with weights for opinion pairs. Parties adjust their policy positions in opinion space to maximize votes, treating votes as potential energy. 
We utilise a square grid for two-dimensional spaces, enabling voters to be maximally extreme on multiple dimensions (i.e. in the corners of the opinion space). This contrasts with a circular model which would mean that voters with extreme positions on one issue were assumed to be moderate on the other.  {\color{black} The two dimensions can be viewed as issues requiring independent legislation, with the bounds representing the Overton Window. }

Our simulation uses a probabilistic Metropolis Monte Carlo method\cite{metropolis-algorithm} which continually explores the favourable regions of opinion space and has a low volatility limit, which is equivalent to finding static minima. 

{\color{black} The implementation is an open-source Python  code which a support easy simulation across $N$-dimensional spaces, facilitating integration of numeric simulations in public choice theory research.}

\subsection*{Voter functions: turnout and activists}\label{model}

Central to solving the vote-maximisation problem, is how each party's votes are calculated: the vote count is the \emph{only} way to influence a party's behaviour in this model. We discuss whether the vote count is fully realistic later, but there is attractive simplicity to the notion that one only needs to change some aspect of the vote calculation in order to model any behaviour one wishes. As is standard, every position  closer to that party than to any other, is counted as `supporting' that party. That region is called the party Voronoi polyhedron (Fig. \ref{fig:voronoi}). In the straightforward Hotelling-Downs model the sum of these positions' voter weights is the party's vote count. For one party in two-dimensional space the vote share is

\begin{equation}\label{base_energy_sum}
    V = \iint_{\text{Voronoi}} W(x, y) \mathrm{d} x \mathrm{d} y = \iint_{\text{Voronoi}} W(\vec{x}) \mathrm{d} \vec{x},
\end{equation}

where $V$ is the total votes and $W(\vec{x})$ is the weight function's value at the coordinate $\vec{x}$, i.e. the number of voters with the pair of opinions $(x, y)$. The integral is over every point in the party's Voronoi polyhedron. Typically, the polyhedra extend to the boundaries of the opinion space: the "Overton Window".

\includegraphics[width=0.9\linewidth]{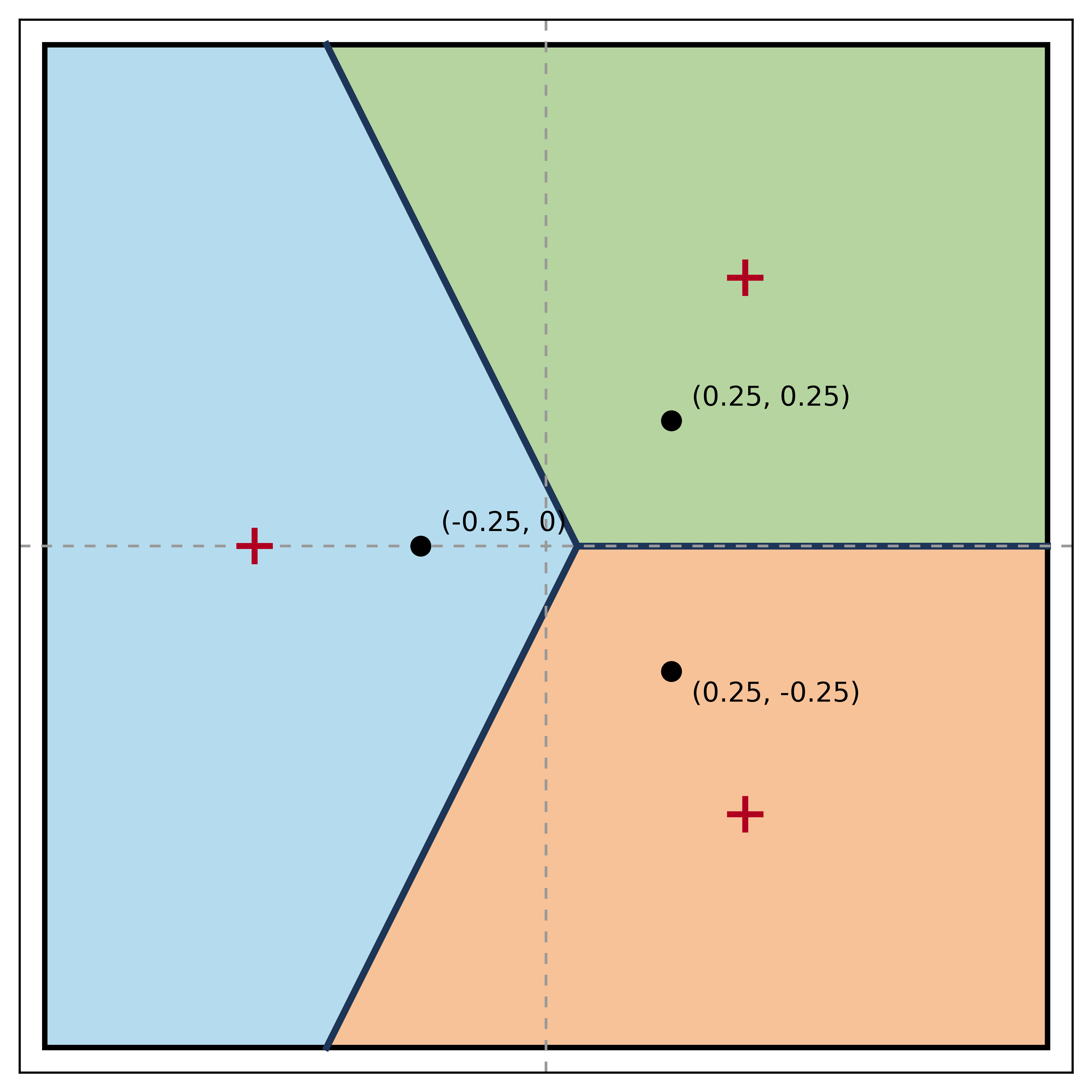}   
\begin{figure}[h]
    \caption{Example of Voronoi polyhedra showing the set of points closer to one party than to any other. Party position is shown by black dots, red crosses show the centroid, i.e. average opinion of that party's voters assuming constant $W(x,y)$. In this example any party can increase its vote share by moving towards the centre, away from the opinion of most of its current supporters.}
    \label{fig:voronoi}
\end{figure}

Equation 1 implies that everyone who supports the party will actually turn out to vote.
In most systems, not all voters will vote, so this integral overestimates the vote. We assume abstention is due to alienation\cite{AdamsMerril2003-TurnoutDiv} so that the probability of abstention increases with the distance between voters' and party's position in opinion space.
We take this function to be an inverse power (with a truncation at probability 1) since it is unlikely that the falloff is linear, but other functions should give similar results. Others, e.g. Callander and Wilson, model turnout as a binary choice but since each opinion position in our model is taken to represent a group of voters, it is more realistic to have the turnout gradually fall to represent individual deviation\cite{Callander&Wilson-Duverger}. For this we introduce the parameter $\tau$. The larger this parameter, the higher the voters' 'propensity to abstain'\cite{Dreyer&Bauer2019}  and the lower the turnout. The antonym  `voter tolerance' has also been used \cite{Callander&Wilson-Duverger} and gives an equivalent model.

Suppressing the multiple integrals, our expression for  the vote count $V_T$ of the party at $\vec{x_p}$ including turnout effects is then

\begin{equation}
    V_T = \int_{\text{Voronoi}} 
        \frac{W(\vec{x})} {(1+(\vec{x} - \vec{x_p})^2)^{\tau}}
    \mathrm{d} \vec{x}.
\end{equation}

This reduces the contribution from voters with opinions far from their nearest party and reverts to Eq. 1 when $\tau\rightarrow 0$.

There may also be an effect favouring extreme positions; more extreme voters may have more clout (see \nameref{discussion}). In that case the vote is weighted according to the radius or polarisation of each voter. Note that mathematically this is equivalent to increasing the voter density at the extremes. Alternatively, it can be viewed as modelling turnout where more extreme voters are more likely to vote, or to be motivated to persuade others to vote. Again we introduce a parameter, $\alpha$, representing their influence and the vote count including these activists becomes,

\begin{equation}
    V_{T,A} =
    \int_{\text{Voronoi}}
    \frac{W(\vec{x})\left|\vec{x}\right|^{2\alpha}}
    {(1+(\vec{x} - \vec{x_p})^2)^\tau}
    \mathrm{d} \vec{x}.
\end{equation}

\subsection*{Polarisation measure}

Previous authors have used various empirical measurements of party polarisation (see citations in \nameref{discussion}). For the purpose of this simulation, the system's party polarisation is defined as the mean historical distance of the parties from the centre. That is, after some equilibration period to account for random initial positions, the radius of each party is averaged over time and then those results are averaged over every party. For example, a party at the centre would have a radius of 0. If it then moved to one of the corners in two-dimensional opinion space (maximum polarisation of 1 in each dimension) it would have a radius of $\sqrt{2}$. To ensure that the simulation is run for sufficiently long that the effect of the initial conditions disappears, the measurement is repeated for different initial conditions and we find that the polarisation is the same, even though the detailed final party positions are not. This measure is simple but sufficient for the purpose of measuring how far the parties in the system deviate from the centre and is appropriate for our highly symmetrical applications.


{\color{black}\subsection*{Voter polarisation despite consensus}

It is an  underappreciated fact that an opinion space can have a single consensus on all issues, but still have multiple maxima.

Suppose the distribution, based on surveys for issues X and Y, is given by the function:  
\[
P(X,Y) \propto \exp\!\left(\frac{-X^{2}+2Y^{2}}{2\sigma^2}\right)\,
\left[\,1 + a\,X\,Y\,\exp\!\left (\frac{-\alpha(X^{2}+2Y^{2})}{2\sigma^2}\right )\right].
\]
with $a<\alpha e/\sigma^2$ to ensure that the distribution function is positive everywhere. 
One can easily verify that the marginal distributions $P(X)$ and $P(Y) $ are Gaussian distributed - for either actual issue, $X$ and $Y$ there is a consensus position, and Y is more sharply peaked.  However, for large a, the correlations mean that  the full distribution $P(X,Y)$ has two distinct maxima away from the origin. 

A simpler example, potentially two-peaked but with single-peaked marginals, is
\begin{equation}
P(X,Y)
\propto 
\exp\!\left(
-\frac{(X-m)^{2} + (Y-m)^{2}}{2\sigma^{2}}
\right) +
\exp\!\left(
-\frac{(X+m)^{2} + (Y+m)^{2}}{2\sigma^{2}}
\right) 
\label{eq:dist} \end{equation}

This is a sum of two 2D Gaussians, with means at $(m,m)$ and $(-m,-m)$.  With $m=0.3$ and $\sigma=1$ they overlap enough to give unimodal marginal distributions (i.e. on each issue there is a consensus position) but the correlation means the distribution as a whole is bimodal.

In each case, the voters are polarised in 2D, but not in either individual issue.  Opinion space weights $W(x,y)$ are set proportional to $P$.}

\subsection*{Model dynamics}
In code, the voter distribution in opinion space is represented by the function value on a fine grid.

To implement the Metropolis algorithm, {\color{black} parties trial new positions displaced from their current position by a vector drawn from a Gaussian distribution (variance 0.02 in the domain [-1,1]), and the change in vote share is calculated.} If a new position yields more votes, it is accepted; otherwise, it is conditionally accepted based on Boltzmann factors: $ P(V_1, V_2) = e^{(V_2 - V_1) / T} $, where $T$ is a volatility constant. This allows for continuous policy evolution and strategic interaction between parties on a grid.  The parties are not located at grid points, thus avoiding ambiguity when they are in precisely the same place.

\section*{Results}\label{results}

\subsection*{Downsian equilibrium}

Our simulations reproduce Hotelling’s Law for both uniform and single‑peaked voter distributions in one‑dimensional opinion space. With two parties, convergence occurs as expected (Table 1).  For three , the system does not possess a Nash equilibrium.  This is because the outer two parties can always gain votes by moving towards the centre, but once they are more than 1/3 from the edge (line length 1) another party can obtain that vote share by "leaping" over it- and one of the other parties must always have less than 1/3 vote share.  A similar argument applies to more parties.

 So a key modelling choice is whether parties may pass one another. Downs argues they should not, because such “leaping” undermines credibility with voters \cite{Downs}. 
 With two parties, Hotelling’s Law holds and the observed polarisation arises purely from Monte‑Carlo volatility.  For three or more parties, the "inside" parties do poorly, so they either leapfrog one of the outer ones, leading to a successive outward motion,  or if passing is forbidden, exert a "pressure" moving the away from the centre.  Any non-leaping move by an inside party gives zero change in vote share: whatever it gains from one neighbour, it loses from the other, so is always accepted by the algorithm.


\begin{figure}[h]
\includegraphics[width=\linewidth]{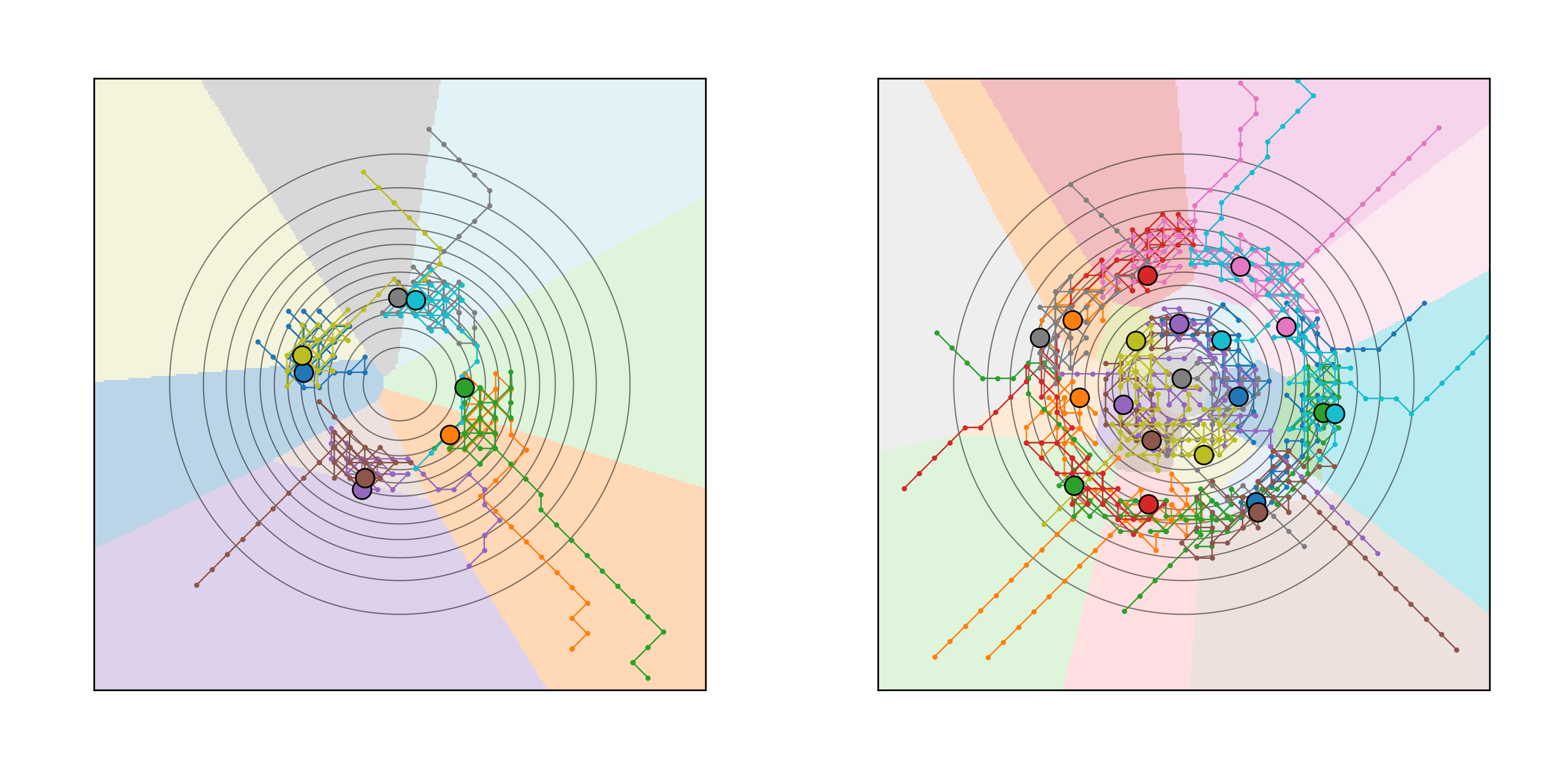}   
    \caption{{\bf Multi party distribution} 
Example equilibria for a unimodal 2D voter distribution ($m=0$ in Eq.\ref{eq:dist}, shown with feint contours), with 8 or 20 parties.  
Lines show typical trajectories from random starting locations, circles show final positions, final Voronoi polyhedra are shaded and final vote-share is stated in the legend.   
    }
    \label{fig:eqm1}
\end{figure}

In the two dimensional case, Hotelling's Law still applies in a two party system, but for more parties our equilibrium search algorithm finds a situation in which parties are located on a ring away from the centre, which ring gets larger as the number of parties increases. For large numbers of parties, a second ring appears (Fig. \ref{fig:eqm1}). 

\begin{figure}[h]
\includegraphics[width=0.9\linewidth]{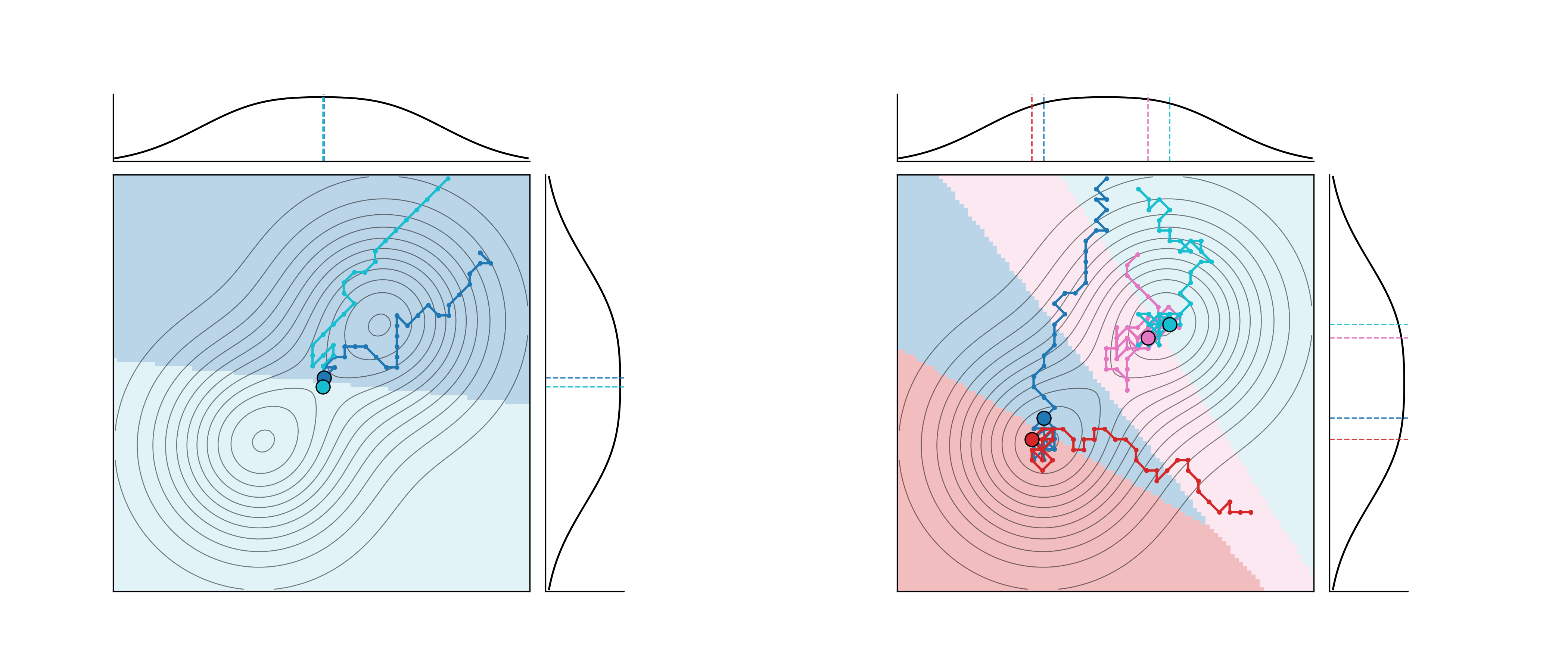}   
    \caption{{\bf Two party consensus and four party polarisation.} 
Equilibria for a bimodal 2D voter distribution, where each issue has a unimodal distribution. Lines show typical trajectories from random starting locations, circles show final positions, final Voronoi polyhedra are as shaded. Marginal distributions as shown, with party positions on each policy.   
    }
    \label{fig:eqm2}
\end{figure}

\begin{table}[h]
    \centering
    \caption{\bf The polarisation and standard error of varying numbers of parties in flat opinion space. The middle column shows values when the parties are allowed to pass each other, the right column when they cannot. Each simulation was run for 2000 steps, the detailed final configurations are all different.  The 2D data is in [-1,1] so is tending towards 0.765 for random placement}.
    \color{black}
    \begin{tabular}{c|c|c|c}
         	Parties &  1D Can Pass & 1D Cannot Pass & Two dimensions\\ \hline
            2	& 0.068(5) &	0.09(1) & 0.02(2)  \\
            3	& 0.34(3)	& 0.43(3)&  0.35(6) \\
            4	& 0.37(2)	& 0.43(5) & 0.46(4)\\
            5	& 0.37(1)	& 0.37(9)  & 0.51(3) \\ 
    \end{tabular}
    \label{tab:1D-Hotelling}
\end{table}{\color{black}
\subsection*{Polarised Voters}
Downs assumed a flat or single-peaked opinion space, and indeed surveys tend to show that on most issues there is a single peaked distribution. However, even if this is true, it is possible to have a bimodal voter distribution because of correlations in opinions.
We now consider the more complex distribution of voters given by the bimodal distribution.

We find the equilibrium positions for the parties numerically with zero-volatility, by discretising the voter distribution on a grid, starting the parties at random positions, and making incremental steps to increase vote share.

The results are shown in Figure \ref{fig:eqm2}.
Remarkably, in a bimodal opinion space, a two-party system goes to the Hotelling solution: both parties adopt the consensus opinion on both issues.  But this efficient representation of voters' wishes fails in a multiparty system: with four parties all four move to the peaks in the voter distribution.
Whichever party "wins" the election in the two party case, the party policy on each issue will correspond to the voters' preference.  By contrast, either coalition of like-minded parties in the four-party case will have policies for both issues away from the voters' consensual preference. In similar runs of the four-party case, initially three parties coalesce in one maximum, then one switched to the other.  Four parties follow Hotelling's law within their peak, six parties split 3-3  and are polarised within the peak, five parties split 2-3, with the pair adopting the Hotelling consensus and the triple being polarised.}

\subsection*{Turnout}
\begin{figure}[h]
\includegraphics[width=0.9\linewidth]{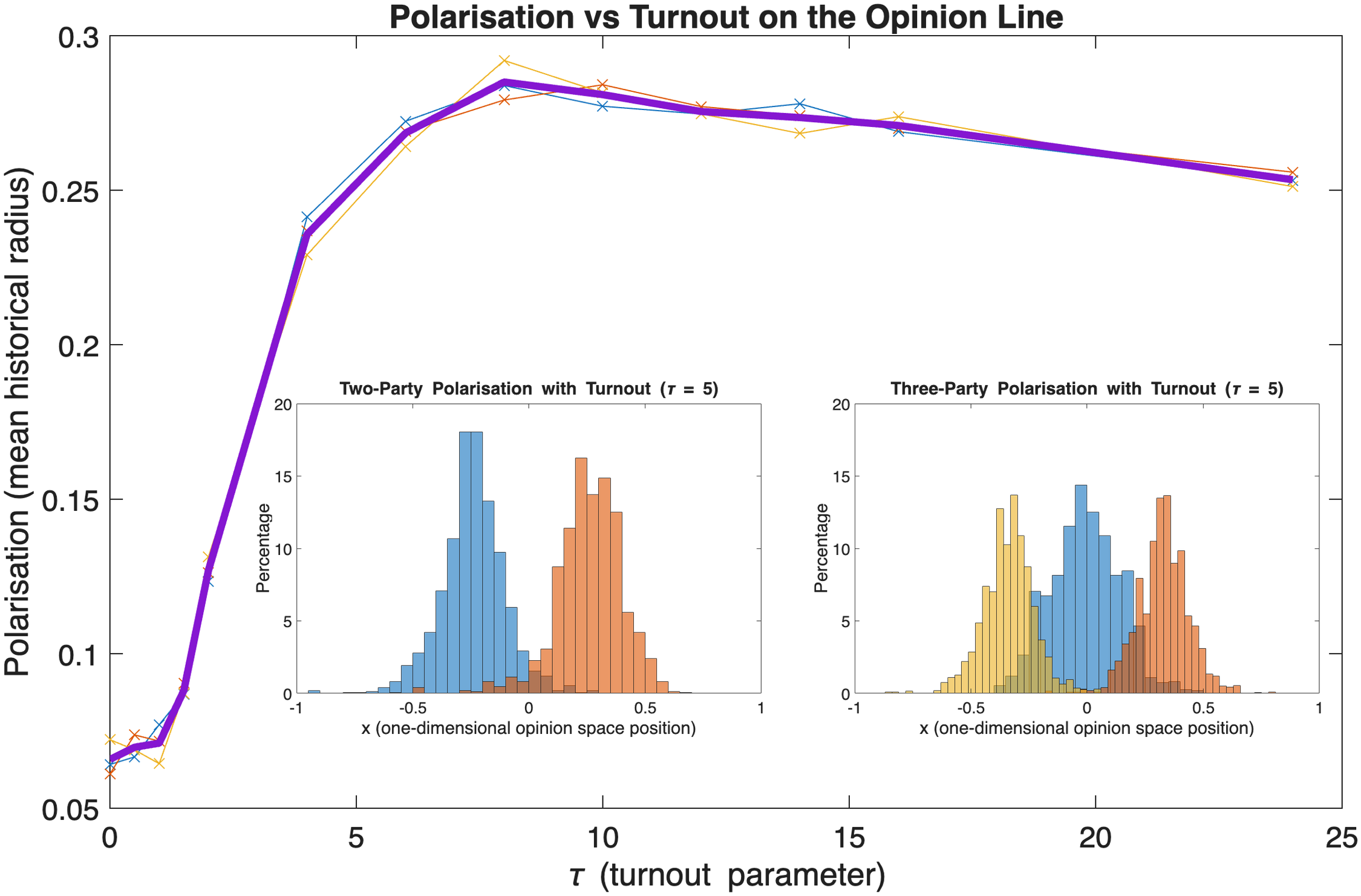}   
    \caption{{\bf Polarisation with turnout} calculated for a single peaked voter distribution  without activists. Polarisation increases with turnout up to a point, while parties are competing directly for votes, but then decreases again when turnout is highly suppressed and centrist voters do not turn out for either party. Inset shows examples for two and three parties.}
    \label{fig:turnout-pol-1D}
\end{figure}

It has been previously reported\cite{Callander&Wilson-Duverger} that turnout effects do not usually significantly increase polarisation for a two party system.  However, we find that this is only true if the voter distribution is strongly peaked in the middle. If the peak is relatively flat or in the limit of the uniform distribution (corresponding to an electorate with more diverse opinions), turnout does lead to polarisation for two or more parties even without the threat of entry of a new challenger (see Fig. \ref{fig:turnout-pol-1D}).
This is because there are votes to be gained from reluctant voters towards the extremes, even at the cost of wavering voters midway between two parties. It is therefore worth polarising a little in order to  better exploit the votes on both flanks of the party, provided the voters do not fall off too quickly towards the extremes. This is consistent with the existence in many countries of strong centre-left and centre-right parties. Both are moderate, but they are sufficiently separated so as not to have too much overlap in their voter bases. 

If this is correct, we should expect polarisation to slightly decrease for extremely high values of $\tau$. This is because the parties separate only so far as to be able to fully exploit the voters available to them. If the spatial distance at which the party can attract a reasonable number of votes is small, the parties may not be competing for the same voters and therefore move closer together since there are still more votes to be found in the centre, \emph{ceteris paribus}. Plotting the effects of turnout on polarisation without activists does indeed show a decrease for very low turnout and high $\tau$ (Fig. \ref{fig:turnout-pol-1D}).

\begin{figure}[h!]
\includegraphics[width=0.9\linewidth]{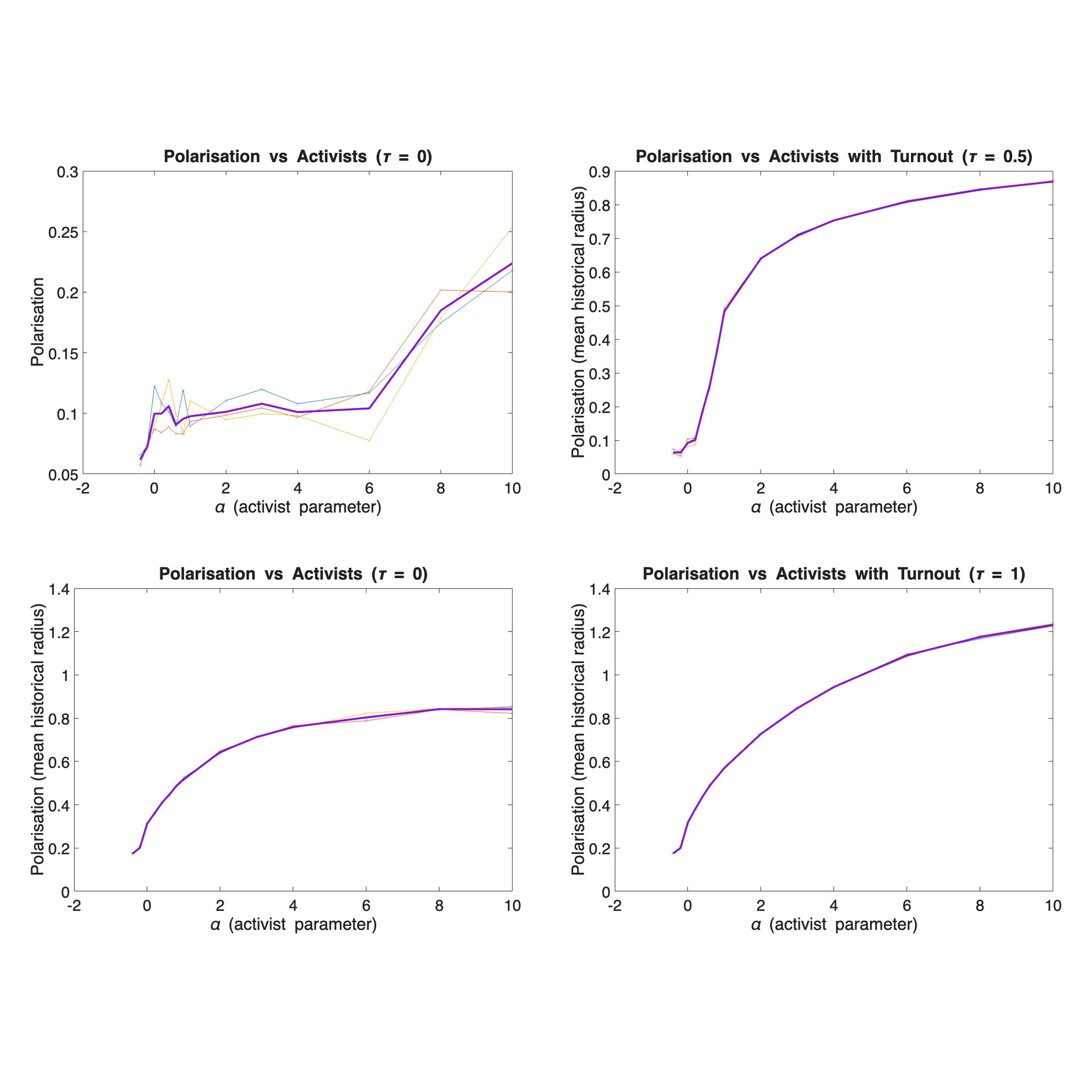}   
    \caption{{\bf Polarisation with activists.} Polarisation in one- and two-dimensional opinion space with a sharp central voter maximum. \emph{Top left:} Activists without turnout have a noticeable but small effect. \emph{Top right: }With turnout activists become more important and polarisation can increase almost to the edge of the line (note the different scale). \emph{Bottom left:} In two-dimensional opinion space, activists alone can have a significant effect. \emph{Bottom right:} Activists with turnout have a much larger effect in two dimensions as well. Each data point comes from running the simulation with the specified $\alpha$ and each point was run three times. Each simulation was run for 2000 steps. The thick purple line tracks the average of the three runs. The fact that the three runs converge to relatively clear curves despite the fact that the parties' initial positions are random indicates that those initial conditions do not have much effect on the outcome so the certainty of these measurements is high.
    }
    \label{fig:params activists+turnout}
\end{figure}

Activists have a polarising effect as expected, since they essentially increase the weighting of the extremes. Interestingly, their effect is much more significant if there is at least a moderate amount of abstention as shown in Fig. \ref{fig:params activists+turnout} (for example $\tau = 0.5$).
The polarisation caused by accounting for activists and turnout together is greater than the sum of its parts. This suggests that multiple factors mat combine non-linearly to explain extreme polarisation. 

In two-dimensional opinion space without turnout or activists,  parties are positioned equidistant from the centre, exploiting wedge-shaped Voronoi polyhedra. Two parties will arrange themselves on opposite sides close to the centre (assuming moderate but opposite positions). Polarisation is low but because the parties have no intrinsic preference for any position, and are subject to random motion from the Monte Carlo process, they rotate around the centre. Three parties polarise,  taking divergent positions and roughly forming an equilateral triangle. The more parties there are, the more the party polarisation increases. We seldom observe parties exchanging policy positions, or moving randomly\cite{Grofman2004}, even though the model parties have no fixed general principles.  

For larger numbers of parties (Fig.\ref{fig:eqm1} we start to see multiple rings of parties around the centre.  The precise number of parties depends on details of $W(x,y)$ but a robust result is that it is symmetric - extreme left and right appear at the same time.

When there is a single ring, the voter base is more extreme than the party.  Party members and activists will be drawn from this voter-base, and will have a pull on the party to move towards the centre of the Voronoi polyhedron.  Activists pull the party towards the edges and turnout towards the centre of their Voronoi polyhedron.
Even for an alternative model where activists are drawn equally from the parties voters, their effect is to pull the party to extremes.

Activists have a fairly large polarising influence on their own, and turnout alone also has a small effect. As in the one-dimensional case, both effects together can produce significant polarisation limited only by the edge of the voter distribution (Fig. \ref{fig:params activists+turnout}). In particular, they can produce an equilibrium.

\subsection*{Dynamic  Equilibrium}

\begin{figure}[h!]
\includegraphics[width=0.9\linewidth]{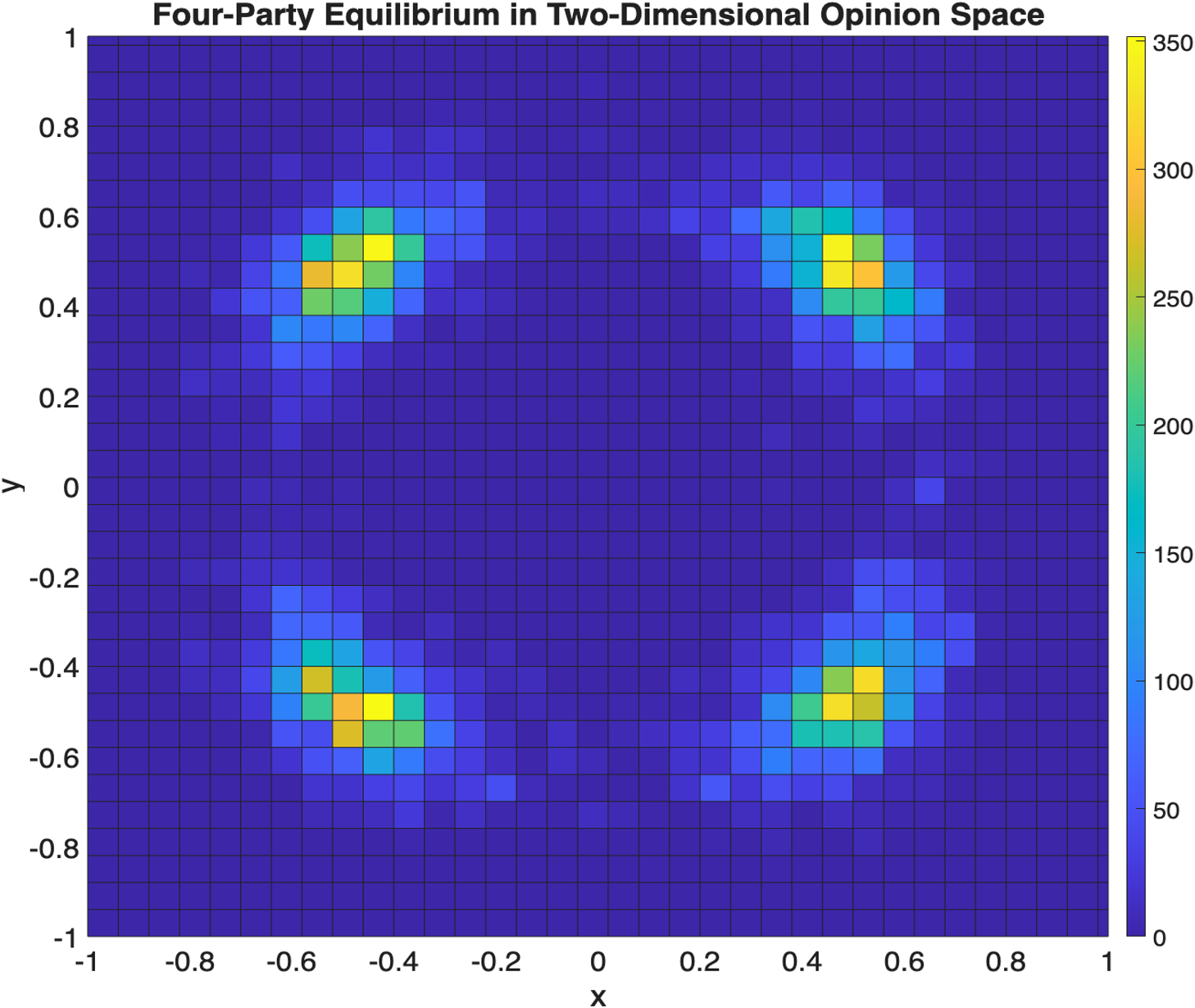}   
    \caption{{\bf Four-party equilibrium.} There exists an equilibrium for four parties in two-dimensional opinion space. Brighter colours indicate a location which the party visited more often, i.e. a more probable and more favourable policy position. It is not favourable for any party to move out of its corner position. This is a balance between the centripetal Hotelling-Downs force, the centrifugal activists ($\alpha=1.5$) and turnout ($\tau=1.5$), which draws parties towards the centre of their electorate. Together these forces create a stable equilibrium which models a polarised party system where each party would lose the support of influential activists if it moved towards the median voter.
    This simulation was run for 5000 steps. The equilibrium is very stable, in the sense that the parties do not switch positions until about 7000 turns for these parameters. It is also independent of initial positions.}
    \label{fig:equilibrium}
\end{figure}

Because our opinion space is a square and not a circle, a symmetric dynamic equilibrium is possible only for certain numbes of parties. It is shown in Fig. \ref{fig:equilibrium}. When activists give polarised voters more weight, the corners become preferable to the edges of opinion space because distance from the centre is higher there. While it does not exist for the Hotelling-Downs case ($\tau=0$ and $\alpha=0$), for higher values of $\alpha \text{ and }\tau$ a system with four parties has a stable equilibrium. When polarisation is low they rotate in a ring near the centre. When it is higher they remain close to symmetric positions  along the diagonals of the space. If a party moved away from its corner it would lose those voters to turnout, and since activists increase with polarisation there are less votes available further inward or along the edges.  
Parties occasionally switch corners due to random fluctuations, but these jumps are extremely rare. Mathematically, this equilibrium exists because by including activists the distribution is no longer uniformly decreasing from the centre.  Politically, one can imagine the radical activists are more motivated to work for a party with policies closer to their opinion.

Asymmetric equilibria can exist for less than four parties as well. When both turnout and activists are present and at least one of those parameters is high, systems with two or three parties will also polarise into the corners. Despite one corner being unoccupied the state can be relatively stable because each party is only affected by the activists near it and the vote cost of moving between corners is very high. This means, for instance, that the model can support a polarised two-party system even in higher-dimensional opinion space. This scenario does provide a clear opportunity for a new party. 


\section*{Discussion}\label{discussion}
With minimal assumptions — namely that parties act to maximise their votes as per Downs, that there is reduced turnout due to abstention and that there are  activists affecting party policy — this Monte Carlo model can reproduce a wide range of party-polarisation systems. We have explicitly shown the one- and two-dimensional opinion space cases, and
extension to $N$-dimensional opinion space is easily possible with minimal recoding. In addition, our model is easy to modify with different postulated forces and combines the theoretical frameworks of statistical mechanics and Downsian economics.

The Hotelling-Downs equilibrium, with all parties colocated centrally adopting more moderate policies than their voters' preference, can exist in our model  with strongly peaked voter distribution. In addition, the two central parties can reach an equilibrium where they are moderately polarised, as is the case for `big tent' centre-left and centre-right parties. This effect is solely due to turnout: the strategy increases party appeal to their base at the expense of losing a few floating voters. Finally, highly polarised equilibria where the party is more extreme than its average voter can exist if the influence of extreme activists is large. Interestingly, the model tends to support symmetric equilibria, so if one party is dragged to the extreme it is likely that the other will become equally extreme in the opposite direction.

Between these cases, our model can reproduce a wide variety of real-world party systems. Many historical and contemporary party systems can be classified as either unpolarised, centred around two moderate parties or strongly polarised. This suggests wide applicability for the model.

Our conclusion supports previous research which claims that single factors alone are not sufficient to explain polarisation (e.g. Callander and Wilson model turnout and the threat of entry of a third party, but find neither force sufficient in isolation), though we show that slight polarisation in mildly peaked systems can result from only turnout considerations\cite{Callander&Wilson-Duverger, AdamsMerril2003-TurnoutDiv}. It is worth noting, however, that the number of factors considered is still relatively small compared to the plethora of factors which might affect a system as complex as voters and political parties (personality of candidates, government record, media coverage, luck, scandal, foreign influence etc.). We do not claim that those factors have no effect, but that it may be possible to combine factors or reduce them to the few most important and still produce a useful model. We will examine the two factors we have chosen more closely below.

Finally, the probabilistic nature of this model is not only useful computationally, but may in fact be a more accurate depiction of reality. Parties do not exactly follow the rational of utility calculations. Parties make mistakes and take risks, but overall they tend to try to maximise their votes. Our stochastic treatment may well be appropriate.

\subsection*{Turnout and internal elections}

The turnout parameter creates a force towards the weighted centre of each party's Voronoi polyhedron. This is because the party wants to minimise abstention due to alienation among its base, and so it is favourable to be closer to as many of its voters as possible. This attractive behaviour towards the party's current electorate can also model internal party elections which also pulls the party policy to the centre of its own Voronoi polyhedron, since only supporters can vote. {\color{black} Typically, the median voter in a party is more extreme than the median voter overall (Fig. \ref{fig:voronoi}), and more extreme than the position adopted by the party under Downsian assumptions.}

Thus, the optimal position in an internal election is in the centre of the party's base. 
The median voter of a party is further explored by Westley \emph{et al.}\cite{median-party-voter}. In particular, closed US presidential primary elections favour more extreme candidates because only party members can vote, while more open primaries favour candidates closer to the overall median voter. Our turnout effect also models this tendency towards the `median voter of the party'\cite{median-party-voter}: an example where different factors lead to the same mathematical representation.

This raises an important point: the votes in the model are not necessarily real votes. They may represent an influence on party decisions which does not translate into votes on election day, or funding which drags many voters to a specific location, creating peaks in $W(x,y)$.
We postulate that every influence on a party can be translated into an effect on its voter potential in the model.
Internal party elections and turnout both have the same effect, drawing the party towards the centre of its supporter base, not of the population as a whole. We model them as one term (associated with $\tau$). Turnout may even indirectly account for the threat of entry of a challenger, since a large group of voters alienated from the existing parties are presumably receptive to a new one. A party interested in maximising its turnout may therefore act similarly to a party worried about a new party threatening it. Since our focus  is primarily on the system as a whole and not on determining winners, whether a party's choice is actually the rational one is not the most important point. Our focus is to quantify the influences on parties' decision making. 

That said, the primary goal of our turnout parameter is to model the effects of abstention, so the vote interpretation is the clearest. The turnout function is taken to be an inverse power, but other decreasing functions should give similar results. Higher powers (higher $\tau$) correspond to lower turnout. Equivalently, each voter's utility of voting falls the further they are from the nearest party.

We have shown that internal elections and reduced turnout can correspond to higher polarisation than the Downsian case. 
In countries with a large share of politically disillusioned people (corresponding to a quick voter falloff when voters slightly disagree with parties), it may be necessary for political parties to assume positions which are  closer to the median voter in their sector of opinion space. We note that without activists, turnout effects move party policy to the centre of the voter base, but not to more extreme positions.

The relationship between turnout and polarization can be confirmed empirically. One can define a number of empirical measures of polarisation, but we adopt that of Van der Veen, who has created a comparative index of polarisation spanning dozens of countries and more than 25 years\cite{Polarisation-Van_der_Veen}. This index accounts for four different types of polarisation, and because of its breadth it allows us to find general trends and draw quite general conclusions, although, because of its generality it may miss specific national effects.  This polarisation is compared with the comprehensive Global Dataset on Turnout (GD-Turnout)\cite{GD-Turnout}. Since we are looking for general trends, we use both the average score for each country in the polarisation index and the average turnout for all elections in the country from GD-Turnout for 1993–2020. This large span of years makes the measurement relatively immune to short-term national fluctuations. We omit countries not present in both datasets.

\begin{figure}[h]
\includegraphics[width=0.9\linewidth]{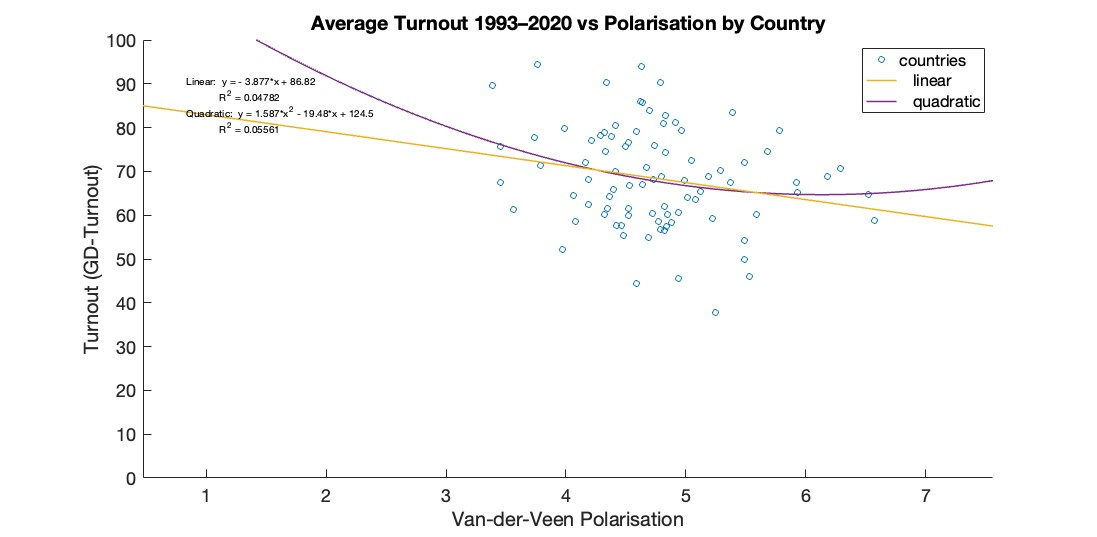}   
    \caption{{\bf Empirical turnout vs polarisation.} Turnout from presidential and legislative elections across 85 countries between 1993 and 2020. Turnout data is taken from GD-Turnout and averaged\cite{GD-Turnout}. It is plotted against polarisation taken from Van-der-Veen\cite{Polarisation-Van_der_Veen}. Despite the noise in the data, there is a statistically significant trend ($p < 0.05$) that decreasing turnout {\color{black} is correlated with}  greater polarisation.}
    \label{fig:real-world}
\end{figure}

Turnout and polarisation across 85 countries between 1993–2020 are plotted in Fig. \ref{fig:real-world}. It shows a weak but significant  correlation between decreasing turnout and increasing polarisation (5\% level (Pearson's $r = -0.22$, $p = 0.0448$).

This comparative analysis lends strong support to our model, which finds turnout moves parties to polarised, but not extreme positions. Not only is turnout empirically correlated with polarisation, but the effect is relatively weak.
Elsewhere in the literature there is significant theoretical and empirical evidence that as turnout declines, polarisation increases\cite{Callander&Wilson-Duverger, Rogowski2014}. How much party polarisation increases for polarised voter distributions (such as our smoothly peaked distribution) is dependent on high propensity to abstain (i.e. high $\tau$, turnout is low)\cite{Dreyer&Bauer2019}. It has been argued based on recent Dutch elections that if there is low turnout, extreme voters may be overrepresented since they are most likely to turn out\cite{yussef2024}. The interplay of turnout and activists increasing extreme votes in our simulation can be understood as precisely this effect.

However, the opposite is also often claimed: increasing polarisation mobilises voters to turn out in greater numbers overall. Hobolt and Hoerner argue that the greater distance between policy outcomes means citizens have more stake in the election\cite{Hobolt2020}. Like this model, they also find that spatial proximity to a party is an important factor in turnout decisions, though they come to the opposite conclusion. In contrast to the above, greater choice thanks to polarisation is also often said to be associated with higher turnout, though some of these results do not seem statistically convincing\cite{Crepaz1990, Wilford2017, Hartveld&Wagner2023}. 

In fact, these opinions may not be opposed.  Our turnout parameter essentially describes the reluctance of voters to turnout at all.  But the movement of the parties is directly aimed at increasing turnout of their base.  So it can be argued that apathy among voters, which directly reduces turnout, leads to parties polarising to increase their voteshare, thereby indirectly increasing overall turnout. 

\subsection*{Activists}

Downs recognised that in the real world, where information is not free and perfect, some people will matter more to the government than others. This population (lobbyists, activists, public speakers, media personalities, representatives, etc.) may not have the same opinion distribution as the population at large, and that is the justification for the activist parameter in our model.
As Downs wrote, if knowledge is imperfect some people can influence others (by giving biased information). `First, it means that some men are more important than others politically, because they can influence more votes than they themselves cast. Since it takes scarce resources to provide information to hesitant citizens, men who command such resources are able to wield more than proportional political influence, \emph{ceteris paribus}'\cite{Downs}. This influence is represented by the increased weighting of activists. Though our model could support centripetal activists for $\alpha<0$ (e.g. charismatic or political power is concentrated among centrists, or even legal sanctions against polices which deviate from the \emph{status quo} too much), we have chosen to analyse them as a centrifugal force $\alpha>0$. This is because the most extreme voters have been found to be the most politically active\cite{median-party-voter, Abramowitz&Saunders2008}. 
The author of that model argues that these more extreme voters influence parties by giving candidates essential support during nomination, helping them during the campaign, and being the most likely source of candidates themselves\cite{activist-model}.
This can include grassroots campaigning, but donors are also generally extreme and their views carry a lot of weight\cite{Levendusky}. 
So the party representatives are also more extreme than vote-winning party policy.

As noted before, even when turnout is low the people most likely to continue going to the voting booths are the most extreme\cite{yussef2024}. In addition, parties in the US have focused significant resources on maximising the turnout of their (more extreme) base\cite{fiorina1999}. Our activists can represent this effect of increased turnout at the extremes.
It therefore makes sense to increase the weight of the extremes, both because they can influence the party in ways other than votes and because, when turnout is considered, the extremes should be more likely to vote.
This is how activists can pull party positions towards them.

Activists can also represent the effect of elite polarisation, if party decision-makers are biased towards extreme policies. Elite polarisation is strongly correlated with other kinds of polarisation, and since there has been a significant increase in it in recent decades, it makes sense to include its effects in our model via extreme activists\cite{BoxellCountryTrends, Fiorina-PolarisationTypes}. It can be concluded that activist-driven polarisation is the result of a small number of people strongly influencing politics by either being more likely to vote or by wielding some sort of elite power.

\section*{Conclusion}\label{conclusion}

We have presented a model with only two parameters but wide applicability. We employ the methods of computational statistical mechanics to implement and enhance a version of Downsian  theory with turnout and activists affecting the voter distribution and the policies of the parties. We have identified equilibria in one and two dimensions wherein no party can benefit by changing its policies.

Our work is intended as a starting point; it is by no means complete. When combining methods from the natural sciences with other disciplines it is beneficial to start with a simple, foundational model to explain general trends and then expand it with researchers from the other discipline and empirical data\cite{BlytheInterdiscipline}. 
Importantly, the Monte Carlo approach makes it quicker and easier to test different theoretical approaches and more complicated multi-peaked, multi-dimensional voter distributions within the Downsian framework because it is relatively easy to modify and recalculate.  

This model's primary omission, common to many Downsian models, is that the voter distribution is static. Nonetheless it demonstrates that it is possible for party polarisation to occur without the electorate's views changing, and despite a plurality of voters being centrists, just as Fiorina and Abrams argue that negligible voter polarisation has occurred while party polarisation has increased in the US\cite{Fiorina-PolarisationTypes}. 

There are also many possible extensions to the basic model which we hope our code can enable.
The exact form of the turnout and activist functions, and empirical values for the parameters in various party systems invite further investigation. In addition, adding or replacing parameters and comparing their results would be very interesting, since it is not self-evident that turnout and activists should be the most important influences on parties. It would also be interesting to allow the number of parties to vary and attempt to reproduce Duverger's law, {\color{black} which would require additional description of the voting system and creation/destruction of parties.  The apparatus for creating and destroying parties exists within analytic statistical mechanics -- the so-called ``Grand Canonical Ensemble". Within the ``opinion space", the voter distribution could be made dynamic, and the boundaries themselves (Overton Window) could move or expand. 
}

{\color{black}
We demonstrate that two-peaked voter polarisation can be created in a 2D space even when each issue has a single-peaked consensus. This happens when there is strong correlation between views on the two issues.

Our main results are to show that Hotelling's Law applies to two-party systems even if the underlying voter distribution is polarised (two peaked).  If there are more parties, they tend  to migrate away from the centre, and for a bimodal distribution they tend to migrate to the peaks. 

Polarisation also appears when voting is optional, or when the parties place the opinions of their supporters ahead of the desire to maximise their vote. 

In both dynamic and static equilibria, the existence of a party with extreme views is strongly correlated with opposite, but equally extreme, views. 

The effect of extra challenger parties at the extreme has a twofold polarising effect.  For example,  a far right party self-evidently adds to polarisation by its very existence, but it also has the effect of drawing the adjacent centre-right party further right\cite{ivaldi2024populist}. 
For high numbers of parties, these may form a ring of strongly polarised parties, including those with single-issue polarised views.
All of this party polarisation may occur despite the existence of consensus among voters on all issues.
 This outcome is exacerbated by abstention and activists.

Movement away from the centre ground already occurs with three-parties to avoid the centremost being "outflanked" by a third party. Interestingly, the result structures tend to have all parties equally polarised, and the more competing parties, the more polarised it becomes.  We sometimes see hotelling-type pairing behaviour within the ring which perhaps provides a mechanism for parties to merge.

 Assuming that an electoral system leads to a government comprising parties receiving a majority of votes,  multi-party induced polarization leads to a more extreme government than a two-party system.  Consequently, although the multiparty system means more voters have a {\it party} and {\it parliament} which represents their views, the {\it government} and {\it government policy} is less representative.
}







\section*{Acknowledgments}
The \texttt{plot\_xyz.py} routine for visualising the output was written by Joe Zuntz, for which we extend our thanks.  CC was supported by an Edinburgh School of Physics career development scholarship.
We thank James Ackland (University of Glasgow) for helpful discussions.
%
%
%

\end{document}